\documentclass[preprint,showpacs,preprintnumbers,amsmath,amssymb]{revtex4}

\usepackage{graphicx}
\usepackage{dcolumn}
\usepackage{amsmath}
\usepackage{amssymb}

\newcommand{\bvec}[1]{\mbox{\boldmath $#1$}}
\newcommand{\D}{\delta }

\def\vq{{\bf q}}

\def\vk{{\bf k}}
\def\vQ{{\bf Q}}

\def\vr{{\bf r}}
\def\vS{{\bf S}}

\newcommand{\eq}[1]{Eq.~(\ref{#1})}
\newcommand{\fig}[1]{Fig.~\ref{#1}}
\newcommand{\be}{\begin{equation}}
\newcommand{\ee}{\end{equation}}
\newcommand{\bea}{\begin{eqnarray}}
\newcommand{\no}{\nonumber}
\newcommand{\eea}{\end{eqnarray}}
\newcommand{\bean}{\begin{eqnarray*}}
\newcommand{\eean}{\end{eqnarray*}}
\newcommand{\bfi}{\begin{figure}}
\newcommand{\efi}{\end{figure}}
\newcommand{\bc}{\begin{center}}
\newcommand{\ec}{\end{center}}
\newcommand{\ba}{\begin{array}}
\newcommand{\ea}{\end{array}}

\begin{document}


\title{Magnetic excitations in La-based cuprate superconductors:   
slave-boson mean-field analysis of the two-dimensional $t$-$J$ model} 

\author{Hiroyuki Yamase} 
\affiliation{Max-Planck-Institute for Solid State Research, 
Heisenbergstrasse 1, D-70569 Stuttgart, Germany}




\date{\today}

\begin{abstract} 
Motivated by recent inelastic neutron scattering experiments 
up to a high-energy region for La-based cuprates, 
we compute $(\vq,\omega)$ maps of the imaginary part of 
the dynamical magnetic susceptibility $\chi(\vq,\omega)$ in the  
slave-boson mean-field approximation to the two-dimensional 
$t$-$J$ model. While the strong spectral weight appears 
at incommensurate positions, namely at $\vq \neq \vQ \equiv (\pi,\pi)$,  
for low energy,  
the incommensurate signals disperse with 
increasing $\omega$ and finally merge into a commensurate signal 
at a particular energy $\omega=\omega_{\vQ}$. 
These features are seen in both the $d$-wave pairing state and 
the normal state. 
In particular, the incommensurate signals below $\omega_{\vQ}$ 
in the normal state are due to the Fermi surface geometry, which we expect 
for La-based cuprates because of a tendency to $d$-wave type 
Fermi surface deformations. 
Above $\omega_{\vQ}$, 
strong signals appear to 
trace an upward dispersion especially for a low doping rate in the 
$d$-wave pairing state while typically broad spectral weight 
is obtained around $\vq=\vQ$ in the normal state. 
Salient features of magnetic excitations in La-based cuprates 
are thus naturally captured in terms of particle-hole excitations.   
Global understanding of magnetic excitations in high-$T_c$ cuprates 
is discussed through a comparison with magnetic excitations in 
YBa$_{2}$Cu$_{3}$O$_{y}$.  
\end{abstract}

\maketitle

\section{Introduction} 
A major theoretical issue to understand the physics of 
high-temperature superconductors concerns their magnetic excitations. 
Most of inelastic neutron scattering measurements 
were performed for two 
high-$T_c$ cuprates, La$_{2-x}$Sr$_{x}$CuO$_{4}$ (LSCO) and 
YBa$_{2}$Cu$_{3}$O$_{y}$ (YBCO). 
For low temperature and low energy, incommensurate (IC) 
magnetic peaks are realized in both LSCO\cite{thurston89,yamada98} and 
YBCO\cite{dai98,mook98,arai99}, 
that is, the peaks of the imaginary 
part of the dynamical magnetic susceptibility $\chi(\vq,\omega)$ 
appear at $\vq =(\pi,\pi\pm 2\pi\eta)$ and $(\pi\pm 2\pi\eta,\pi)$; 
$\eta$ parametrizes the degree of incommensurability. 
Above the superconducting transition temperature $T_{c}$ or 
possibly above the pseudogap temperature $T^{*}$,  
the IC peaks merge into a commensurate peak 
in YBCO\cite{dai98,arai99,bourges00} 
while LSCO still retains IC signals up to a temperature much higher than 
$T_c$.\cite{thurston89,aeppli97,ito03,fujita04} 

The size of the incommensurability $\eta$ 
depends on the excitation energy $\omega$. 
For YBCO, $\eta$ decreases with increasing $\omega$ and vanishes
at a specific energy $\omega_{\vQ}^{\rm res}$.
This commensurate signal was called a 
``resonance peak'',\cite{rossatmignod91,bourges98,fong00,dai01} 
which is now regarded to be continuously connected with 
the IC signals observed for lower energy, that is the peak disperses 
smoothly downwards to lower energy when $\vq$ is shifted away from 
$\vQ$.\cite{arai99,bourges00,reznik04,pailhes04} 
Above the resonance energy, on the other hand, 
peaks of magnetic excitations 
trace an upward dispersion.\cite{arai99,hayden04} 
For LSCO, the energy dependence of $\eta$ was 
relatively weak compared to YBCO\cite{mason92} and 
such a robustness of the IC signals was often contrasted 
with the behavior in YBCO. 
However, recent high energy neutron scattering data for 
LSCO with $x=0.10$ and $0.16$,\cite{christensen04} 
and La$_{2-x}$Ba$_{x}$CuO$_{4}$ (LBCO) with $x=0.125$\cite{tranquada04} 
revealed that the IC peaks disperse with $\omega$. 
In particular, it was found in LBCO that the IC peaks merge into a 
commensurate peak around 55 meV,\cite{tranquada04} above which 
upward dispersive features appear, similar to YBCO 
except that the data for YBCO were obtained below $T_c$ while 
those for LBCO above $T_c$. 

Magnetic excitations in La-based cuprates are often interpreted 
in terms of the spin-charge stripe scenario,\cite{tranquada95} 
according to which (a tendency to) charge stripe ordering 
plays a central role. 
This scenario is based on the observation of two charge order satellite 
signals at both $\vq=(0,\pm 4\pi\eta)$ and $(\pm 4\pi\eta,0)$, whose 
wavevectors are just twice as large as those of the magnetic IC 
signals,  in Nd-doped LSCO with $x=0.10$,\cite{ichikawa00} $0.12$,\cite{tranquada95,zimmermann98} and $0.15$,\cite{niemoller99} 
La$_{2-x}$Ba$_{x}$CuO$_{4}$ (LBCO) with $x=0.125$,\cite{fujita04}  
and Sr-doped LBCO with the hole density $0.125$.\cite{fujita02b,kimura03} 
It is assumed that each CuO$_2$ plane has charge stripes 
characterized by a single wavevector 
$\vq=(0,\pm 4\pi\eta)$ or $(\pm 4\pi\eta,0)$. 
Hence the observed two charge order signals are speculated to be 
superposition originating from different CuO$_2$ planes. 
Although such charge order signals were very weak and were 
observed neither for other doping 
rates nor in other cuprate superconductors such as YBCO and LSCO, 
the stripe scenario has attracted much interest.\cite{kivelson03} 

There is another scenario based on conventional particle-hole 
excitations around the Fermi 
surface (FS), often refereed to as the fermiology 
scenario.\cite{si93,zha93,tanamoto93,tanamoto94,liu95} 
This scenario was explored in more detail and turned out to 
capture many important features of magnetic excitations 
in both YBCO\cite{bulut96,millis96,brinckmann99,kao00,norman00,manske01,chubukov01,brinckmann02,onufrieva02,li02,schnyder04,sega0306,yamase06} and LSCO.\cite{kao00,manske01,yamase010203} 
In particular, to be reconciled with a Fermi surface shape implied by 
angle-resolved photoemission spectroscopy 
(ARPES),\cite{ino9902,zhou01} the fermiology scenario for LSCO 
was extended by invoking the idea of a $d$-wave type Fermi surface 
deformation ($d$FSD),\cite{yamase00} which is frequently called 
$d$-wave Pomeranchuk instability:\cite{pomeranchuk58} 
the FS expands along the $k_{x}$ direction and shrinks along the 
$k_{y}$ direction, or vice versa. 
Comprehensive calculations of magnetic excitations for 
this $d$-wave-deformed FS\cite{yamase010203} 
showed that 
prominent features of magnetic excitations in LSCO are 
well-captured within the fermiology scenario. 
However, these calculations were confined to a 
low energy region of magnetic excitations. 
It is interesting to see whether the fermiology scenario can capture 
recently reported data for a high energy region in La-based 
cuprates also,\cite{tranquada04} since such data were often discussed 
within spin-charge stripe scenarios.\cite{vojta04,uhrig04,seibold05}

In this paper, we extend a previous calculation in Ref.\cite{yamase010203}, 
which was intended for magnetic excitations in the low energy region 
for LSCO, to the high energy region. 
We compute the dynamical magnetic susceptibility $\chi(\vq,\omega)$ 
in the slave-boson mean-field scheme of the two-dimensional (2D) 
$t$-$J$ model. Combining the idea of the $d$FSD with magnetic excitations, 
we show that salient features of magnetic excitations in La-based 
cuprates are well-captured up to high energy 
in terms of particle-hole excitations. We discuss how 
similarities and differences of magnetic excitations between LSCO and YBCO 
arise in the present theory.

The paper is structured as follows. 
In Sec.~II, we present the slave-boson mean-field scheme of 
the $t$-$J$ model and show how the $d$FSD is obtained in this scheme. 
The dynamical magnetic susceptibility $\chi(\vq,\omega)$ 
is formulated in the RPA with 
a renormalization factor.\cite{brinckmann99,yamase99} 
In Sec.~III, numerical results of $(\vq,\omega)$ maps of 
Im$\chi(\vq,\omega)$ are presented up to high energy for several choices of 
temperature $T$ and hole density. 
In Sec.~IV, we compare our results with neutron scattering data in 
La-based cuprates and discuss similarities and differences of magnetic 
excitations between La-based cuprates and YBCO from the present 
theoretical viewpoint. Sec.~V is a conclusion.

\section{Model and formalism}

We take the 2D $t$-$J$ model on a square lattice 
\be
 H = -  \sum_{\vr,\,\vr',\, \sigma} t^{(l)} 
 \tilde{c}_{\vr\,\sigma}^{\dagger}\tilde{c}_{\vr'\,\sigma}+
   J \sum_{\langle \vr,\vr' \rangle} \,
 \vS_{\vr} \cdot \vS_{\vr'}  \label{tJ} 
\ee  
defined in the Fock space with no doubly occupied sites.
The operator $\tilde{c}_{\vr\,\sigma}^{\dagger}$
($\tilde{c}_{\vr\,\sigma}$) creates (annihilates) an electron with
spin $\sigma$ on site $\vr$, while $\vS_{\vr}$ is the 
spin operator. 
$J(>0)$ is a superexchange coupling between the 
nearest neighbor sites. 
We take into account hopping amplitudes $t^{(l)}$ 
up to the $l$th nearest neighbors $(l\leq 3)$, which are denoted 
by the conventional notation, $t$, $t'$, and $t''$, respectively.  

We introduce the slave particles, $f_{\vr \sigma}$ and $b_{\vr}$, 
as $\tilde{c}_{\vr \sigma}=b_{\vr}^{\dagger}f_{\vr \sigma}$,  
where $f_{\vr \sigma}$ ($b_{\vr}$) is a fermion (boson) operator 
that carries spin $\sigma$ (charge $e$), and $\vS_{\vr}=\frac{1}{2}
f_{\vr \alpha}^{\dagger}{\bvec \sigma}_{\alpha \beta} f_{\vr \beta}$ 
with the Pauli matrices ${\bvec \sigma} = (\sigma^x,\sigma^y,\sigma^z)$.
The slave bosons and fermions are linked by the local constraint
$b_{\vr}^{\dagger} b_{\vr} + 
 \sum_{\sigma} f_{\vr \sigma}^{\dagger} f_{\vr \sigma} = 1$.
This is an exact transformation known as the slave-boson formalism. 
We then decouple the interaction with the so-called 
resonating-valence-bond (RVB) mean fields:\cite{fukuyama98} 
$\chi_{\boldsymbol \tau}$$\equiv$$\langle \sum_{\sigma}f_{\vr\,\sigma}^{\dagger}
f_{\vr' \,\sigma}\rangle$, 
$\langle b_{\vr}^{\dagger}b_{\vr'}\rangle$, and 
$\Delta_{\boldsymbol \tau}$$\equiv$$\langle f_{\vr\,\uparrow}f_{\vr' \,\downarrow}- 
f_{\vr\,\downarrow}f_{\vr' \,\uparrow}\rangle$, 
with $\bvec{\tau}=\vr'-\vr$. 
These mean fields are assumed to be real constants independent of sites $\vr$. 
We approximate the bosons to condense at the bottom of the band, which 
leads to 
$\langle b_{\vr}^{\dagger}b_{\vr'}\rangle=\D$, where $\D$ is the hole 
density. 
The resulting Hamiltonian reads 
\be
\hspace{-0mm}H_{0}=\sum_{\vk}
\left(
      f_{\vk\,\uparrow}^{\dagger}\;\; f_{-\vk\,\downarrow}
\right)
\left( \begin{array}{cc} 
   \xi_{\vk} & -\Delta_{\vk} \\
-\Delta_{\vk} & -\xi_{\vk}
          \end{array}\right)
\left( \begin{array}{c}
 f_{\vk\,\uparrow} \\
 f_{-\vk\,\downarrow}^{\dagger}
\end{array}\right)\, \label{MFH}
\ee
with a global constraint 
$\sum_{\sigma}\langle f^{\dagger}_{\vr\sigma}f_{\vr\sigma}\rangle =1-\D$; 
the $\vk$ summation is over $|k_{x(y)}| \leq \pi$. 
The RVB 
mean fields enter in $\xi_{\vk}$ and $\Delta_{\vk}$, and  are determined 
by minimizing the free energy. We obtain $\chi_{x}=\chi_{y}=\chi$ 
and $\Delta_{x}=-\Delta_{y}=\Delta$ ($d$-wave paring). 
The explicit momentum dependence of the dispersion is given by 
\bea
&&\hspace{-0mm}\xi_{\vk}=-2\left[ 
  \bar{t}_x \cos k_{x} + \bar{t}_y\cos k_{y}
 \right.\no\\ 
&& \hspace{-3mm}+ \left. 2t'\D \cos k_{x} \cos k_{y}
+ t''\D \left(\cos 2k_{x}+\cos 2k_{y}\right)\right]- \mu \,, 
\label{xi}
\eea
where $\bar{t}_x=\bar{t}_y=\bar{t}=t\D+\frac{3}{8}J \chi$ and 
$\mu$ is the chemical potential, and 
the gap function is given by 
\be
\hspace{-0mm}\Delta_{\vk}=-\frac{3}{4}J\Delta   
 \left(\cos k_{x}-\cos k_{y}\right)
\label{sc} \,. \\
\ee 

The material dependence of high-$T_{c}$ cuprates 
is taken into account mainly by different choices of band 
parameters.\cite{si93,tanamoto93,feiner96,tohyama00,pavarini01} 
Following previous work for La-based 
cuprates,\cite{tanamoto93,yamase010203} 
we choose $t/J=4,t'/t=-1/6$, and $t''/t=0$. 
In the course of the present work, we noticed that 
incommensurate structures 
of Im$\chi(\vq,\,\omega)$ in the normal state become clearer for a 
larger $t/J$, showing better agreement with neutron scattering data. 
Note that the present choice of $t/J$ is consistent with an 
{\it ab initio} calculation\cite{hybertsen90}, which showed 
that realistic values for $t/J$ lie in the range $2-5$.  

The obtained FS is electronlike (holelike) 
for $\D \gtrsim 0.10$ ($\D \lesssim 0.10$) [\fig{FS}(a)]. 
This FS seems to be inconsistent with ARPES data 
where a holelike FS was implied in a much wider hole-doping region 
($\D \lesssim 0.20$);\cite{ino9902}  the reported FS 
looked nearly the same as a FS in YBCO.\cite{schabel98} 
If this were the case, 
a fermiology scenario predicts that magnetic excitations in 
LSCO should be essentially the same as those in YBCO, which 
contradicts with neutron scattering data especially for the normal state. 
This problem was considered in the 2D $t$-$J$ model. 
It was found 
that the system has a tendency to a $d$-wave type Fermi 
surface symmetry breaking.\cite{yamase00}  
The same tendency was also found in the Hubbard model.\cite{metzner00} 
This $d$FSD tendency is generated by forward scattering interactions 
of electrons close to the FS around $(\pi,0)$ and $(0,\pi)$, and 
may be a generic feature as seen in 
various works: the slave-boson 
mean-field theory,\cite{yamase00,yamase04b}   
the exact diagonalization analysis,\cite{miyanaga06} 
and the variational Monte Carlo study\cite{edegger07} of the $t$-$J$ model, 
various renormalization group studies\cite{metzner00,wegner02,honerkamp02} 
and the renormalized perturbation theory\cite{neumayr03} of the 
Hubbard model, and the mean-field theory of the extended 
Hubbard model.\cite{valenzuela01}

\begin{figure}
\centerline{\includegraphics[width=0.45\textwidth]{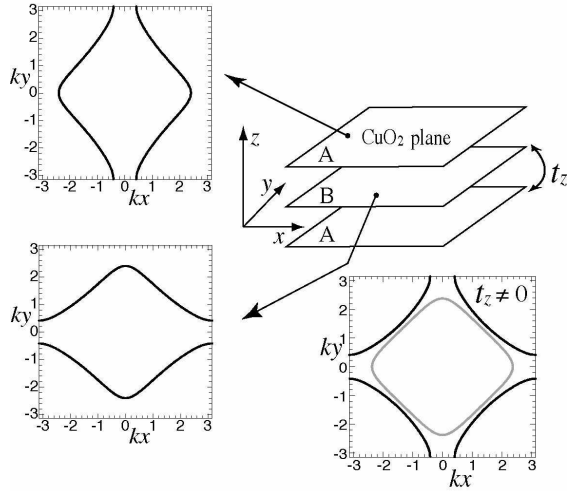}}
\caption{Alternate stacking of $d$-wave deformed Fermi surfaces 
along the $z$ axis in the presence of the LTT in La-based cuprates. 
Because of a weak interlayer coupling $t_{z}$,  
the resulting Fermi surfaces (right panel) 
are composed of a holelike 
Fermi surface (solid line) and an electronlike Fermi surface (gray line).} 
\label{alternateFS}
\end{figure}

The $d$FSD competes with $d$-wave superconductivity.\cite{yamase00,metzner00,honerkamp02,wegner02,neumayr03,yamase04b,edegger07}  
In the $t$-$J$ model,\cite{yamase00,yamase04b,edegger07}
the $d$-wave pairing instability is dominant and the spontaneous 
$d$FSD does not happen. However, the system still retains appreciable 
correlations of the $d$FSD,\cite{yamase00,yamase04b,miyanaga06,edegger07} 
which make the system very sensitive 
to a small external anisotropy, 
leading to a strongly deformed FS. 
A natural origin of this anisotropy lies in crystal structures. 
La-based cuprates have the low-temperature tetragonal lattice (LTT) 
structure.\cite{axe89,crawford91,sakita99} 
It yields a small $xy$ anisotropy, the direction of which 
alternates along the $z$-axis. 
Through a coupling to the LTT, therefore, 
we can expect an alternate stacking of 
a strongly deformed FS 
as shown in \fig{alternateFS}. 
Because of a weak interlayer coupling, the resulting Fermi surfaces 
have two sheets, an inner electronlike FS and an outer holelike FS. 
The FS reported by Ino {\it et al.}\cite{ino9902}  may correspond to the 
outer holelike FS; the 
inner electronlike FS has not been detected successfully by ARPES.

Although the deformed FS can be determined in a self-consistent 
calculation,\cite{yamase00,yamase06}  we here 
deform a  FS by hand by introducing  a parameter $\alpha$ 
\be
\bar{t}_{x(y)} = \alpha \bar{t}
\label{a}
\ee
with keeping $\bar{t}_{y(x)}= \bar{t}$. 
For simplicity we do not consider an interlayer coupling 
and compute magnetic excitations for the superimposed FSs shown 
in \fig{FS}(b). We first determine RVB mean fields 
self-consistently for $\alpha=1$, and then tune $\alpha$ to get a 
FS compatible with ARPES;\cite{ino9902} we choose $\alpha=0.85$ 
for $\D=0.15$ [\fig{FS}(b)].  
Since we here tune $\alpha$ by hand, the resulting hole density 
slightly deviates from $\D=0.15$ to be $0.156$.

We investigate magnetic excitations for both FSs shown in 
Figs.~\ref{FS}(a) and (b). 
The irreducible dynamical magnetic susceptibility 
$\chi_{0}(\vq,\,\omega)$ is given by 
\bea
\chi_{0}(\vq,\,\omega) &=& \frac{1}{4 N} 
\sum_{\vk}\left[C^{+}_{\vk,\,\vk+\vq} 
 \left(\tanh \frac{E_{\vk}}{2T}
   -\tanh \frac{E_{\vk +\vq}}{2T}\right)
   \frac{1}{E_{\vk}-E_{\vk+\vq}+\omega+{\rm i}\Gamma}\right. \no\\
 &+& \frac{1}{2} \, C^{-}_{\vk,\,\vk+\vq}
\left(\tanh \frac{E_{\vk}}{2T}
   +\tanh \frac{E_{\vk +\vq}}{2T}\right) \no\\
 && \times \left. \left(\frac{1}{E_{\vk}+E_{\vk+\vq}+\omega+{\rm i}\Gamma}
   +\frac{1}{E_{\vk}+E_{\vk+\vq}-\omega-{\rm i}\Gamma}\right)\right],\,
\label{xoqw}
\eea
where $E_{\vk}=\sqrt{\xi_{\vk}^2+\Delta_{\vk}^2}$, 
$\Gamma$ is a positive infinitesimal, and 
\be
C^{\pm}_{\vk,\,\vk+\vq}=\frac{1}{2}
\left(1 \pm \frac{\xi_{\vk}\xi_{\vk+\vq}
 +\Delta_{\vk}\Delta_{\vk+\vq}}{E_{\vk}E_{\vk+\vq}}\right)\; .
\label{coherence}
\ee
In a renormalized random phase approximation 
(RPA)\cite{brinckmann99,yamase99} 
the dynamical magnetic susceptibility $\chi(\vq,\,\omega)$ is given by 
\be
\chi(\vq,\,\omega)=\frac{\chi_{0}(\vq,\,\omega)}{1+ 
J(\vq)\chi_{0}(\vq,\,\omega)}\; , \label{RPA}
\ee
where 
\be
J(\vq) = 2r J(\cos q_{x}+\cos q_{y})  \label{RPAJ}
\ee
with a renormalization factor $r$.
In the 
plain RPA one has $r=1$, which leads to a divergence of $\chi(\vq,\,0)$ 
around $\vq \sim (\pi,\pi)$ in a wide hole-doping region 
$(\delta \lesssim 0.17)$ in the $d$-wave pairing state, 
signaling an instability to the antiferromagnetic (AF) state. 
However, fluctuations not included in the RPA 
obviously suppress the AF instability as shown in several numerical 
studies.\cite{chen90,giamarchi91,himeda99} 
This aspect may be roughly taken into account in a phenomenological way by 
setting $r<1$.\cite{brinckmann99,yamase99} 
Here we choose $r = 0.35$, which confines the AF instability 
to $\delta \lesssim 0.02$, consistent with the phase diagram 
of LSCO.\cite{keimer92}

\section{Results}
\begin{figure}
\centerline{\includegraphics[width=0.45\textwidth]{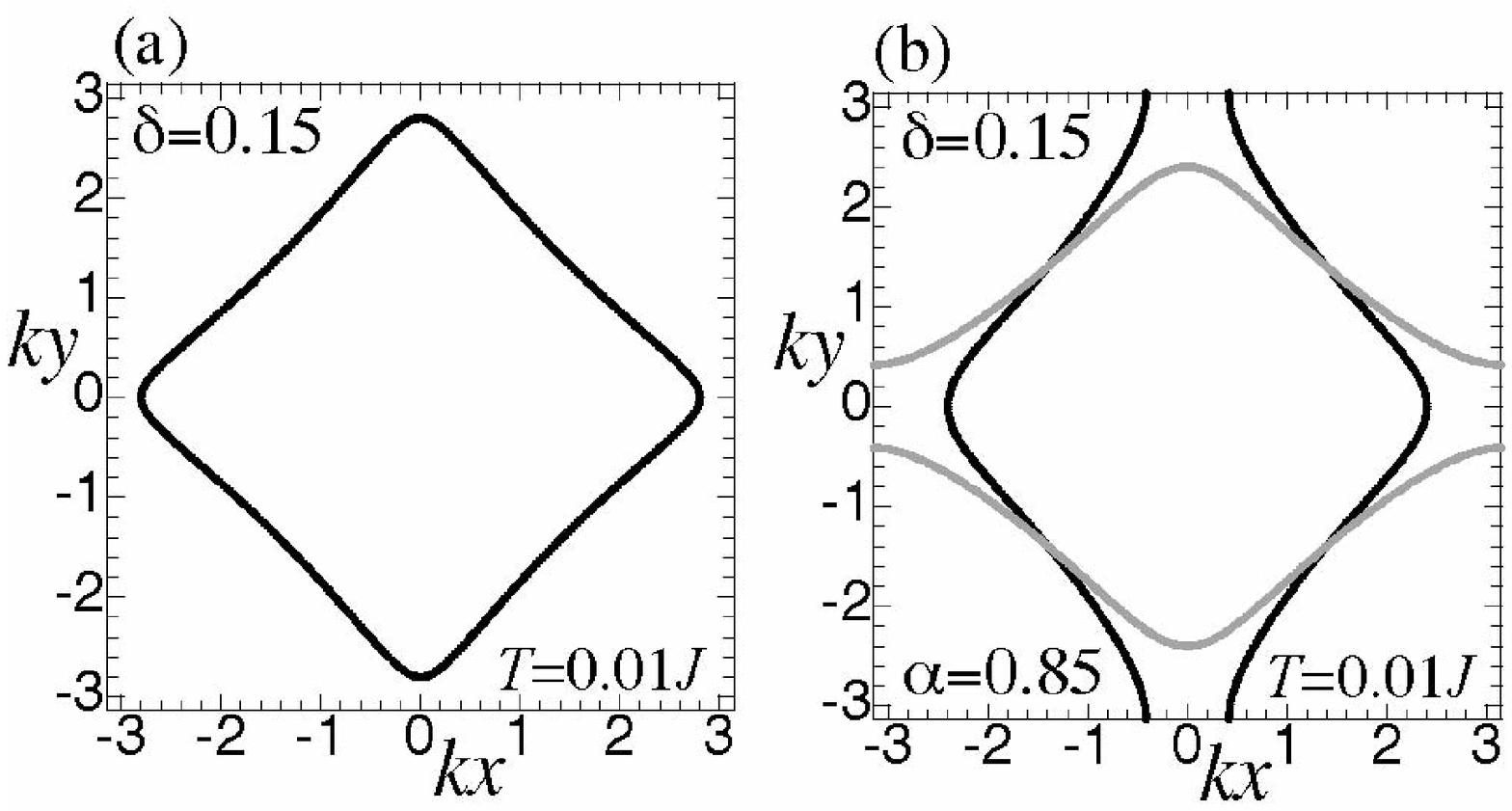}}
\caption{Fermi surface for $\D=0.15$: (a) isotropic case 
and (b) anisotropic case with $\alpha=0.85$, where two Fermi surfaces 
(solid line and gray line) are superimposed. The Fermi surface is 
defined by $\xi_{\vk}=0$ at $T=0.01J$ in the $d$-wave pairing state.}  
\label{FS}
\end{figure}

We compute $\chi(\vq,\omega)$ numerically for the FS shown in \fig{FS}(a) 
and then for that shown in \fig{FS}(b). 
A positive infinitesimal $\Gamma$ in \eq{xoqw} is replaced 
by $\Gamma=0.01J$ in the 
$d$-wave pairing state and $\Gamma=0.05J$ in the normal state. 
Although the choice of a finite $\Gamma$ is done for numerical convenience, 
it may simulate damping of quasiparticles by static defects in 
real materials and broadening due to limited energy resolution in 
inelastic neutron scattering experiments.  

\begin{figure}
\centerline{\includegraphics[width=0.8\textwidth]{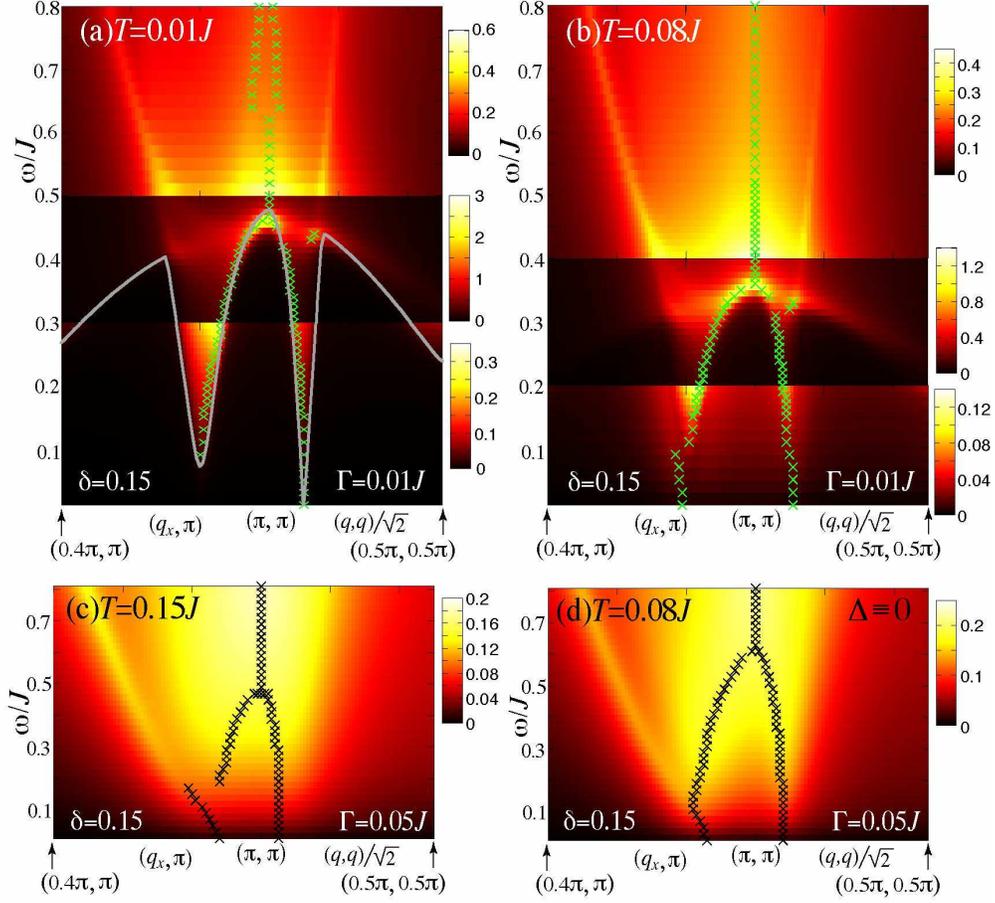}}
\caption{(color online) 
$(\vq,\omega)$ maps of Im$\chi(\vq,\omega)$ for $\D=0.15$ at  
several choices of $T$ for the Fermi surface shown in \fig{FS}(a); 
$T_{RVB}=0.104J$, but $\Delta\equiv 0$ is assumed in (d); 
the $\vq$ scan directions are $(0.4\pi,\pi) \leq \vq \leq (\pi,\pi)$ 
and $(\pi,\pi) \geq \vq \geq (0.5\pi,0.5\pi)$; the cross symbols represent 
the highest weight positions along $\vq=(q_x,\pi)$ and $(q,q)/\sqrt{2}$; 
the gray line in (a) is a lower edge of the continuum excitations; 
(a) and (b) are made of three different energy regions with an 
optimal color map scale.  
}
\label{qwt-L1}
\end{figure}

Figure~\ref{qwt-L1} shows $(\vq, \omega)$ maps of the imaginary part of 
$\chi(\vq,\omega)$ for the FS shown in \fig{FS}(a) for several 
choices of temperatures; $\vq$ is scanned 
along $(0.4\pi,\pi) \leq \vq \leq (\pi,\pi)$ and 
$(\pi,\pi)\geq \vq \geq (0.5\pi,0.5\pi)$; along each direction 
the highest weight positions of Im$\chi(\vq,\omega)$ are represented by 
cross symbols; in \fig{qwt-L1}(a) we also plot 
the threshold energy of individual particle-hole excitations by 
a gray line; Figures~\ref{qwt-L1}(a) and (b) are composed 
of three different energy regions with an optimal color map scale. 
There appear gapless excitations 
along $\vq=(q,q)/\sqrt{2}$ in \fig{qwt-L1}(a). 
This diagonal IC signal is 
due to scattering processes between gap nodes of 
the $d$-wave singlet pairing, 
as already clarified about a decade ago.\cite{zha93,tanamoto94}  
With increasing $\omega$, scattering processes around IC positions 
at $\vq=(\pi\pm 2\pi\eta,\pi)$ and $(\pi,\pi\pm 2\pi\eta)$ 
begin to contribute. The highest spectral weight positions 
(cross symbols) appear close to the threshold energy for both the IC 
and the diagonal IC signals; the IC signal becomes stronger 
than the diagonal IC signal. 
The incommensurability $\eta$ typically tends to decreases with  $\omega$, 
resulting in a commensurate peak at a particular energy 
$(\omega\approx 0.45J)$, which we define as $\omega_{\vQ}$; 
note that $\omega_{\vQ}$ 
corresponds also to a peak energy of Im$\chi(\vQ,\omega)$ in the 
$d$-wave pairing state. 
Looking closely at a region near $\omega_{\vQ}$, we find that 
the peak position of Im$\chi(\vq,\omega)$ is located slightly 
below the threshold energy, where the denominator in \eq{RPA} vanishes, 
that is, Im$\chi(\vq,\omega)$ has a pole. 
Hence the strong signals near 
$\omega \approx \omega_{\vQ}$ are interpreted as collective particle-hole 
excitations, namely the so-called resonance. 
But the peak position appears very close to the threshold energy, 
not well inside the gap. 
Thermal broadening and the spectral weight broadening due to 
a finite $\Gamma$ [\eq{xoqw}]  
easily make the resonance overdamped through 
mixing with individual particle-hole excitations. 
In addition, the $(\vq,\omega)$ region where the resonance is realized 
is very limited. 
Therefore, particle-hole excitations below $\omega_{\vQ}$ are 
mainly characterized by a downward dispersion of individual excitations, 
which traces the $\vq$ dependent threshold energy. 

Above $\omega_{\vQ}$, the strong spectral weight appears around 
$\vq=\vQ$ (cross symbols). We can, however, read off 
from \fig{qwt-L1}(a) that such a commensurate signal is accompanied by 
IC and  diagonal IC substructures for 
$\omega \gtrsim 0.5J$, 
which disperse outwards with increasing $\omega$. 

The $(\vq, \omega)$ map in \fig{qwt-L1}(a) is robust 
against temperature as seen in \fig{qwt-L1}(b) 
and the downward dispersion below $\omega_{\vQ}$ 
remains even in the normal state, namely above 
$T_{\rm RVB}(=0.104J)$ [\fig{qwt-L1}(c)]; 
$T_{\rm RVB}$ is the onset temperature of the $d$-wave pairing gap, and 
is interpreted as the pseudogap crossover temperature $T^{*}$ 
in the underdoped regime 
and as the superconducting phase transition temperature $T_{c}$ 
in the overdoped regime of high-$T_{c}$ cuprates.\cite{fukuyama98}  
Although IC and diagonal IC peaks are seen in \fig{qwt-L1}(c), 
the peak structures are very broad because of 
thermal broadening. 
When temperature is reduced under the condition $\Delta\equiv 0$ 
($\chi_{\tau}$ and $\mu$ are determined self-consistently), 
we see clearer IC and diagonal IC signals in \fig{qwt-L1}(d); 
the former is stronger than the latter. 
These incommensurate structures in the normal state 
can be traced back to the FS geometry shown in \fig{FS}(a). 
Besides the presence of incommensurate nesting vectors, 
there are no particle-hole scattering 
processes across the FS with $\vq=\vQ$ and a small $\omega$, 
which substantially reduces the spectral weight around $\vQ$, leading  
to robust incommensurate structures of magnetic excitations. 
In Figs.~\ref{qwt-L1}(c) and (d), 
we see a IC substructure above $\omega \sim 0.2J$ along $\vq=(q_{x},\pi)$, 
which disperses outwards with increasing $\omega$. 
This substructure becomes stronger for a larger $\D$ 
and a smaller $\Gamma$.

\begin{figure}
\centerline{\includegraphics[width=0.8\textwidth]{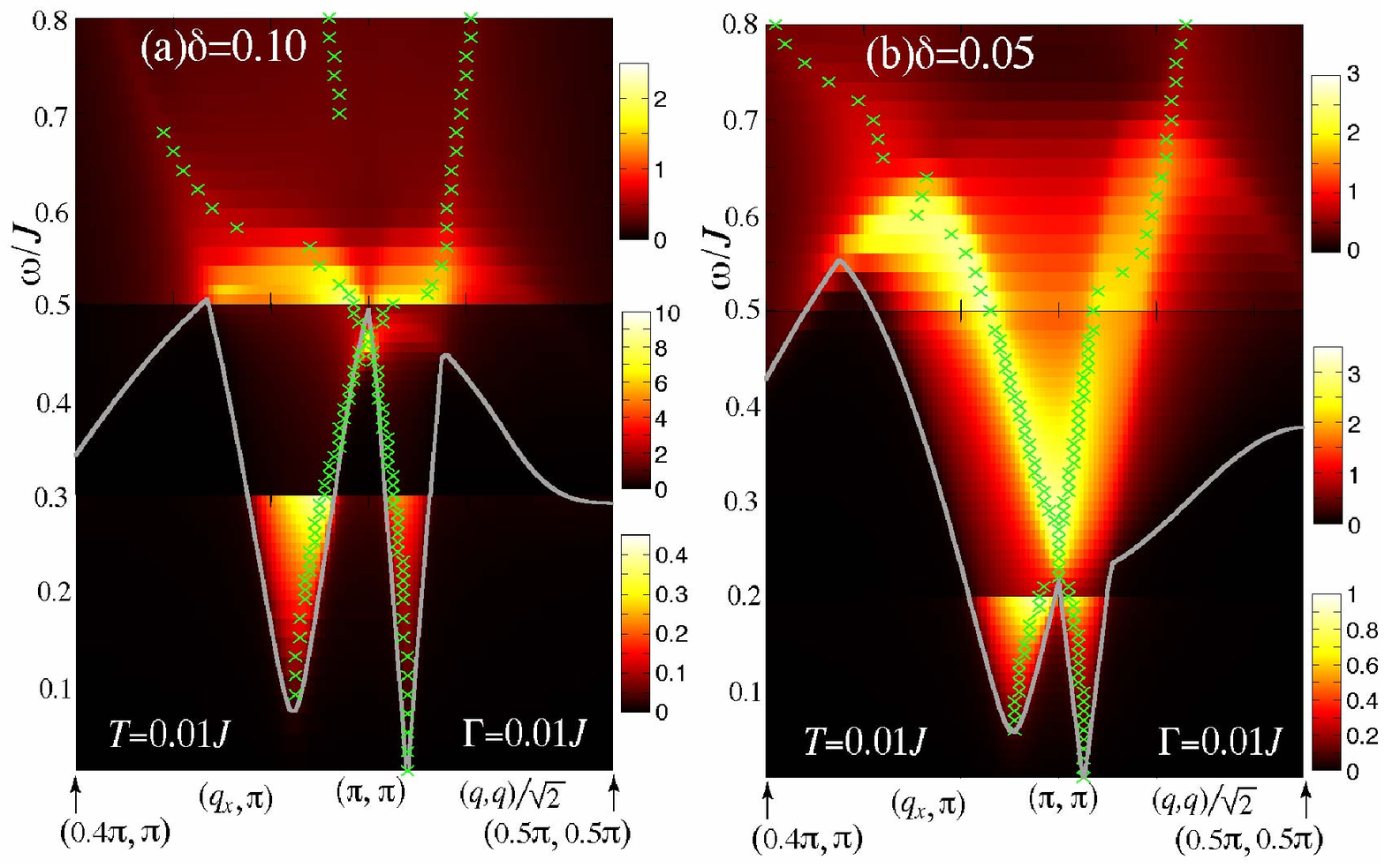}}
\caption{(color online) 
$(\vq,\omega)$ maps of Im$\chi(\vq,\omega)$ at $T=0.01J$ 
for $\D=0.10$ (a) and 0.05(b); 
the corresponding result for $\D=0.15$ is shown in \fig{qwt-L1}(a). 
}
\label{qwd-L1}
\end{figure}

Next we analyze the $\D$ dependence of Im$\chi(\vq,\omega)$. 
Figure~\ref{qwd-L1} shows  $(\vq,\omega)$ maps of Im$\chi(\vq,\omega)$ 
in the $d$-wave pairing state ($T=0.01J$). 
A downward dispersive feature  
below $\omega_{\vQ}$ still appears for a lower $\D$, accompanied by 
a reduction of $\omega_{\vQ}$. 
Above $\omega_{\vQ}$, on the other hand, an upward dispersive feature  
shows up with decreasing $\D$.  
In particular, its spectral weight becomes larger than that of the 
downward dispersion in \fig{qwd-L1}(b) (see values of color map index). 
This upward dispersion is not due to collective excitations, but just 
a peak of individual particle-hole excitations. 
In the vicinity of the AF instability $(\D \approx 0.02)$, 
however, this upward dispersion changes into 
an overdamped collective mode 
in the sense that the real part of the denominator of \eq{RPA} vanishes. 
In the normal state, low-energy incommensurate structures become 
less clear for a lower $\D$ since the FS geometry tends to allow 
particle-hole scattering processes with $\vq=\vQ$ on the FS. 

\begin{figure}
\centerline{\includegraphics[width=0.8\textwidth]{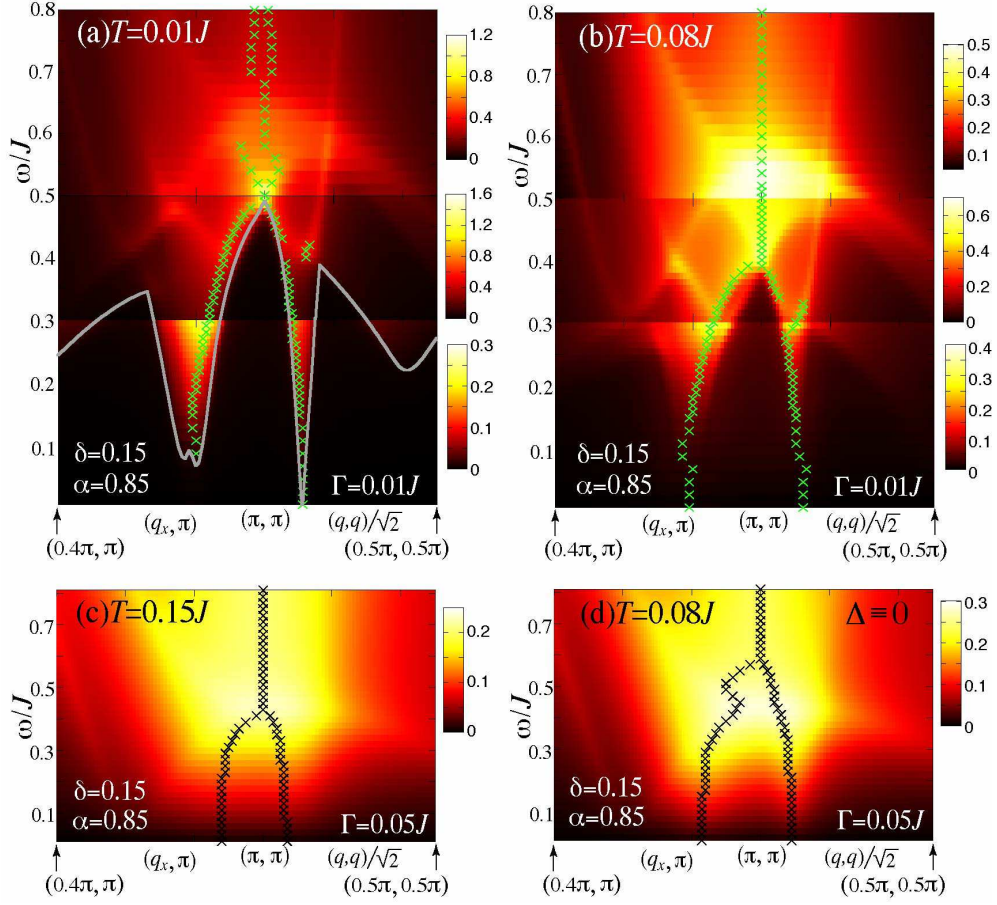}}
\caption{(color online) 
$(\vq,\omega)$ maps of Im$\chi(\vq,\omega)$ for $\D=0.15$ at 
several choices of $T$ for the Fermi surface shown in \fig{FS}(b); 
$T_{RVB}=0.104J$, but $\Delta\equiv 0$ is assumed in (d); 
the corresponding results for the Fermi surface in \fig{FS}(a) 
are shown in \fig{qwt-L1}. 
}
\label{qwt-L2}
\end{figure}

Results similar to Figs.~\ref{qwt-L1} and \ref{qwd-L1} 
are obtained for the FS shown in 
\fig{FS}(b). 
$(\vq, \omega)$ maps of Im$\chi(\vq,\omega)$ are shown in 
\fig{qwt-L2} for several choices of $T$. 
In \fig{qwt-L2}(a), peak positions of  Im$\chi(\vq,\omega)$ (cross symbols) 
trace a downward dispersion below $\omega_{\vQ} (=0.49J)$, which 
comes from individual particle-hole excitations. 
While a collective mode, namely the resonance, does not appear 
in \fig{qwt-L2}(a) ($\D =0.15$), it can appear for a lower doping region. 
But even in this case, we checked that the collective mode is not well 
separated from the continuum excitations 
and thus easily overdamped for a finite 
$T$ and $\Gamma$. Moreover the collective mode appears in a very limited 
$(\vq,\omega)$ region as seen in \fig{qwt-L1}(a). 
In this sense, the downward dispersion below $\omega_{\vQ}$ is 
characterized mainly by individual excitations even for a 
lower doping rate. 
Although spectral weight above 
$\omega_{\vQ}$ is characterized by a commensurate signal at 
$\vq=\vQ$, we see a small segment of an upward dispersion just above 
$\omega_{\vQ}$ in \fig{qwt-L2}(a). 
When the hole density is decreased, this upward dispersion 
becomes clearer and a result similar to \fig{qwd-L1} is obtained.   
Overall features seen in \fig{qwt-L2}(a) still survive 
for a higher temperature [\fig{qwt-L2}(b)],  
although the temperature in \fig{qwt-L2}(b) is high enough to smear 
the segment of an upward dispersion 
for $\omega \gtrsim \omega_{\vQ}$. 
Even above $T_{\rm RVB} (=0.104J)$, incommensurate structures are still 
seen in \fig{qwt-L2}(c) and 
become clearer when temperature is reduced under the condition of 
$\Delta\equiv 0$ [\fig{qwt-L2}(d)]. 
The robust property of the incommensurate structures in the 
normal state [Figs.~\ref{qwt-L2}(c) and (d)] can be 
understood in terms of the geometry of each deformed FS 
shown in \fig{FS}(b) by the same argument as that 
already given in the context of \fig{qwt-L1}. 
In reality, there is a weak interlayer coupling and thus the resulting FSs 
are composed of a holelike FS and an electronlike FS 
as shown in \fig{alternateFS}, which then 
allows low-energy scattering processes with $\vq=\vQ$ through 
the interlayer coupling. Such effects are, however, too weak to smear 
incommensurate structures for a realistic parameter 
as was explicitely calculated in Ref.~\cite{yamase010203}.

\section{Discussion}

\subsection{Magnetic excitations in La-based cuprates}

Now we compare our results with neutron scattering data. 
For both FSs shown in Figs.~\ref{FS}(a) and (b), we have obtained 
IC magnetic excitations for a relatively low $\omega$ 
not only in the $d$-wave pairing state but also in the normal state 
[Figs.~\ref{qwt-L1} and \ref{qwt-L2}], which well captures the most 
prominent feature of magnetic excitations in La-based 
cuprates.\cite{thurston89,aeppli97,ito03,fujita04}  

The incommensurability typically decreases with increasing $\omega$ and 
the signals finally merge into a commensurate peak 
at $\omega=\omega_{\vQ}$. This dispersive feature is consistent 
with recent data for LBCO with $x=0.125$,\cite{tranquada04} 
and also for LSCO with $x=0.10$ and $0.16$\cite{christensen04} where 
measurements were performed up to a certain energy below $\omega_{\vQ}$.  

For a higher energy region in LBCO with $x=0.125$, 
Tranquada {\it et al.}\cite{tranquada04} reported upward dispersive 
signals of Im$\chi(\vq,\omega)$. 
Their obtained spectral weight distribution was very broad with 
substantial spectral weight at $\vq=\vQ$, and the intensity difference 
between IC (or diagonal IC) positions and a commensurate position 
was very small. 
Our results for $\D=0.15$ (Figs.~\ref{qwt-L1} and \ref{qwt-L2}) 
show typically 
a broad spectral weight around $\vq=\vQ$ for a high 
energy region, but an upward dispersive feature shows up with decreasing 
$\D$ in the $d$-wave pairing state (\fig{qwd-L1}). 
Although their measurement\cite{tranquada04} was performed above $T_{c}$, 
we may assume the data were  
obtained in the pseudogap state in LBCO, which is associated with 
the $d$-wave pairing state in the slave-boson 
mean-field theory.\cite{fukuyama98} 
Under this assumption,  
we can capture their data within the present study.

Because of $d$FSD correlations, the shape of the FS 
in La-based cuprates may depend strongly on the crystal 
structure.\cite{yamase00} 
In the presence of the LTT structure, which was 
observed in LBCO\cite{axe89}, Nd-doped LSCO\cite{crawford91}, and 
LSCO with $x=0.12$\cite{sakita99}, we expect that $d$FSD correlations 
lead to a strongly deformed FS as shown in \fig{alternateFS} 
through coupling to a small $xy$ anisotropy of the lattice. 
The corresponding magnetic excitations are shown in \fig{qwt-L2}. 
Even in the absence of the (static) LTT, 
the soft phonon mode toward the LTT structural phase transition, 
whose energy $\omega_{\rm ph}$ 
is about a few meV, was observed in LSCO with 
$x\lesssim 0.18$\cite{thurston89b,lee96,kimura00,wakimoto04} 
and LBCO with $x=0.125$\cite{kimura05}. 
In this case we expect dynamical fluctuations of the $d$FSD 
within a time scale shorter than $\omega_{\rm ph}^{-1}$. 
Thus high energy probes such as ARPES and neutron scattering 
may see an instantaneous $d$FSD. 
On the other hand, in the absence of the LTT and its slow fluctuations, 
which we expect roughly for 
$T \gtrsim 100-200$K,\cite{thurston89b,lee96,kimura00,wakimoto04,kimura05} 
no driving force leading to a strongly 
deformed FS may be present. We expect a FS as in \fig{FS} (a) 
and magnetic excitations as in Figs.~\ref{qwt-L1} and \ref{qwd-L1}. 
In the present $d$FSD correlation scenario, therefore,
the change of the FS shape is predicted as function of $T$. 
The ARPES data for high $T$ have not yet been obtained for 
La-based cuprates and this prediction can be tested in a future.

An indirect evidence of the FS change was recently obtained in a 
neutron scattering experiment for LBCO with $x=0.125$. 
This material shows almost a first-order-like 
LTT structural phase transition at 60K.\cite{kimura05} 
Fujita {\it et al.}\cite{fujita04} measured a temperature dependence 
of the incommensurability $\eta$ at a low energy and found 
a sizable change of $\eta$ at the LTT transition. 
This is naturally understood in terms of 
the (static) FS change scenario from \fig{FS}(a) to \fig{FS}(b) below 60K; 
the FS shown in \fig{FS}(b) tends to favor a larger $\eta$, the 
same tendency as the experiment. 
The authors in Ref.~\cite{fujita04}, however, interpreted the data 
differently 
as a lock-in effect that the periodicity of charge stripes tends to be 
commensurate with the lattice potential of the LTT structure, 
although a charge order signal developed gradually  below 50K, 
not directly below the LTT transition temperature. 

The state with a $d$-wave deformed FS has the same symmetry 
as the so-called electronic nematic state.\cite{kivelson98} 
The nematic order was extensively discussed for cuprates.\cite{kivelson03}  
However this nematic order was discussed as coming from 
partial spin-charge stripe order, not from $d$FSD correlations; 
magnetic excitations were then discussed in terms of 
stripes. 
Our $d$FSD does not require charge stripes nor their fluctuations, 
but  is driven by forward scattering processes of quasiparticles, which 
provides another route to the nematic state. 
An interesting open question is whether the $d$FSD state might 
have an instability 
toward a charge ordered state such as stripes. 
Even if it were the case, our calculation shows that 
many prominent features of magnetic excitations 
observed in La-based cuprates 
are already well-captured without stripes. 
Observations of weak charge order signals\cite{ichikawa00,tranquada95,zimmermann98,niemoller99,fujita04,fujita02b,kimura03} 
do not necessarily mean that charge stripes are crucial to magnetic 
excitations. 
Effects of a charge order on magnetic excitations can be higher 
order corrections beyond the present renormalized RPA and might 
be responsible for a realization of static IC antiferromagnetic order 
around hole density 1/8 in La-based cuprates,\cite{tranquada95,suzuki98,kimura99,ichikawa00,fujita02b,fujita04} 
which cannot be captured in the present framework.

\subsection{Comparison with magnetic excitations in YBCO} 
We finally discuss how similarities and differences of magnetic 
excitations between La-based cuprates and YBCO arise from 
the present framework. 
In the slave-boson mean-field theory, material dependences are 
described by different choices of band parameters. 
For La-based cuprates, we have taken $t/J=4$, $t'/t=-1/6$, $t''/t=0$, 
and $r=0.35$ (see Sec.~II) while 
$t/J=2.5$, $t'/t=-0.30$, $t''/t=0.15$, and $r=0.5$ were invoked for YBCO 
(see Ref.~\cite{yamase06}, where the bilayer coupling 
was also taken into account and a comprehensive study of magnetic 
excitations in YBCO was performed including effects of the $d$FSD). 

A well-known distinction between La-based cuprates and YBCO is that 
low-energy IC magnetic signals are realized 
up to a temperature much higher than $T_c$ in the former,\cite{thurston89,aeppli97,ito03,fujita04} while they are realized only below $T_{c}$ or 
possibly below the 
pseudogap temperature $T^{*}$ in the latter.\cite{dai98,arai99,bourges00}   
This difference comes from a FS difference due to different choices of 
$t'/t$ and $t''/t$. 
For the FS shown in \fig{FS}, there are no 
particle-hole scattering processes with $\vq=\vQ$ for low energy, 
yielding robust incommensurate structures even in the normal state, 
while the FS for YBCO\cite{schabel98} 
allows such scattering processes, 
smearing incommensurate features in the normal state. 
We confirm early works\cite{si93,zha93,tanamoto93,tanamoto94} 
to understand this material dependence of magnetic excitations. 

In the $d$-wave superconducting state, on the other hand,  
both LSCO\cite{yamada98,christensen04} and YBCO\cite{arai99,bourges00,reznik04,pailhes04} show IC magnetic signals for low $\omega$ 
and its incommensurability $\eta$ decreases with increasing $\omega$. 
This behavior is well-reproduced in the present framework, and 
does not depend on the choice of band parameters as emphasized 
before.\cite{kao00}  

The so-called resonance signals  of magnetic excitations 
were reported for YBCO in the 
superconducting state\cite{rossatmignod91,bourges98,fong00} 
and a pseudogap state\cite{dai01}, 
but not for LSCO. 
In the present theory, this resonance is interpreted as 
collective particle-hole excitation as already discussed by many 
authors.\cite{liu95,bulut96,millis96,brinckmann99,kao00,norman00,manske01,chubukov01,brinckmann02,onufrieva02,li02,sega0306,schnyder04,yamase06} 
The realization of collective excitations depends strongly 
on the choice of $r$ in \eq{RPAJ}; it is more favorable for a larger $r$. 
Although our relatively small $r$ for LSCO can produce a resonance, 
its energy appears very close to the threshold energy of continuum 
excitations, making the resonance easily overdamped by thermal 
broadening and spectral broadening due to a finite $\Gamma$ [\eq{xoqw}]. 
Therefore the resonance signals 
are not so clear for LSCO as for YBCO. It is to be noted that if 
we invoke a larger $r$ for LSCO, keeping other parameters unchanged, 
well-defined resonance signals appear as in the case of the parameters  
for YBCO.\cite{yamase06} 
There is another factor, a FS difference, 
which is less effective than a factor $r$. 
Since the FS invoked for LSCO (\fig{FS}) is closer to the 
$(\pi/2,\pi/2)$ point 
than that for YBCO, lowest-energy particle-hole 
scattering processes with $\vq=\vQ$ appear at $k_x=k_y$ 
when a large $d$-wave paring gap opens on the FS. 
For such scattering processes, the coherence factor $C^{-}_{\vk, \vk+\vq}$ 
[\eq{coherence}] vanishes. Hence at zero temperature the imaginary part of 
$\chi_{0}(\vQ,\omega)$ does not show a jump at the threshold energy, 
but increases continuously; 
the corresponding real part of $\chi_{0}(\vQ,\omega)$ does not have 
a log-divergence there. 
This is the case for $\D <0.08$ for the present parameters for 
La-based cuprates. 
Since one of well-known mechanisms of the resonance\cite{liu95,bulut96,millis96,brinckmann99,kao00,norman00,manske01,chubukov01,brinckmann02,onufrieva02,li02,sega0306,schnyder04,yamase06} is based on this log-divergence of 
Re$\chi_{0}(\vQ,\omega)$, 
the resonance is further disfavored for La-based cuprates.

Above the resonance energy, the so-called upward dispersive mode was found 
in YBCO\cite{hayden04,reznik04,pailhes04}, which was captured 
in the present slave-boson mean-field framework\cite{yamase06}  
as well as other phenomenological calculations.\cite{kao00,norman00} 
A similar high energy spectrum, which was much broader than that in YBCO, 
was also reported for LBCO with $x=0.125$.\cite{tranquada04} 
Our results in Figs.~\ref{qwt-L1},\ref{qwd-L1}, and \ref{qwt-L2} capture 
this experimental data. 
The realization of an upward dispersion becomes more 
favorable for a smaller $\D$, a larger $r$, and a smaller $t/J$. 
Since the latter two factors are not respected 
for the present parameters for La-based cuprates, 
the origin of the upward dispersion is different from that for  YBCO:
a peak of individual particle-hole excitations 
(except for the vicinity of the AF instability) for La-based cuprates and 
overdamped collective excitations\cite{yamase06} for YBCO. 

\section{Conclusion}

We have computed Im$\chi(\vq,\omega)$ up to a high energy 
region ($\lesssim J$) and shown that salient features of 
magnetic excitations in La-based cuprates are well-captured 
in terms of particle-hole excitations. 
IC magnetic signals are realized for low $\omega$ in both the 
$d$-wave pairing state and the normal state, and 
they disperse with increasing $\omega$, merging into a 
commensurate signal at $\omega=\omega_{\vQ}$. 
The resonance mode can be realized 
close to $\omega=\omega_{\vQ}$,  
which is easily overdamped by mixing with individual particle-hole  
excitations for finite $T$ and $\Gamma$. 
Above $\omega_{\vQ}$, upward dispersive features of 
Im$\chi(\vq,\omega)$ appear especially for a low $\D$ in the $d$-wave 
pairing state,    
while typically broad spectral weight distribution is obtained 
around $\vq=\vQ$ in the normal state. 
In the present theory, $d$FSD correlations\cite{yamase00,metzner00} 
have played an important role for the shape of the FS. 
In particular, we expect a change of a (static) FS shape 
across the LTT structural phase transition, which in general 
yields a sizable change of incommensurability. 
Such a change was recently observed in LBCO.\cite{fujita04}  
Combined with previous work\cite{yamase010203} focusing on the low energy 
region of magnetic excitations, the present study confirms the crucial 
role of particle-hole excitations up to high energy 
for magnetic excitations in La-based cuprates.

\begin{acknowledgments} 
The author is grateful to C. Bernhard, M. Fujita, 
and R. Zeyher for helpful discussions, and especially to W. Metzner 
for a critical reading the manuscript. 
\end{acknowledgments}


\bibliography{main.bib}

\newpage

\end{document}